\documentclass[twocolumn]{aa}  
\usepackage{graphicx}
\usepackage{txfonts}
\newcommand{\etal}{et al.}

\def\arcm{\hbox{$^\prime$}}
\def\arcs{\hbox{$^{\prime\hskip -0.1em\prime}$}}

\begin{document}

   \title{Cross-Calibration of the XMM-Newton EPIC pn \& MOS On-Axis Effective
     Areas Using 2XMM Sources}

   \author{A.\,M. Read\inst{1},
           M. Guainazzi\inst{2},
           S. Sembay \inst{1}}

   \offprints{A.\,M. Read}

   \institute{\inst{1}Dept.\ of Physics and Astronomy, Leicester University, Leicester LE1\,7RH, U.K.\\
              \email{amr30@star.le.ac.uk} \\
             \inst{2}XMM-Newton Science Operations Centre, ESAC, Apartado 78, 28691 Villanueva de la Ca\~{n}ada, Madrid, Spain\\
}

   \titlerunning{Cross-Calibration of the XMM-Newton pn \& MOS On-Axis Effective
     Areas }

   \date{Received September 15, 1996; accepted March 16, 1997}


  \abstract
{}
{We aim to examine the relative cross-calibration accuracy of the on-axis effective
  areas of the XMM-Newton EPIC pn and MOS instruments.}
{Spectra from a sample of 46 bright, high-count, non-piled-up isolated on-axis
  point
  sources are stacked together, and model residuals are examined to
  characterize the EPIC MOS-to-pn inter-calibration. }
{The MOS1-to-pn and MOS2-to-pn results are broadly very similar. The cameras
  show the closest agreement below 1\,keV, with MOS excesses over pn of 0-2\%
  (MOS1/pn) and 0-3\% (MOS2/pn). Above 3\,keV, the MOS/pn ratio is consistent
  with energy-independent (or only mildly increasing) excesses of 7-8\%
  (MOS1/pn) and 5-8\% (MOS2/pn). In addition, between 1-2\,keV there is a
  `silicon bump' $-$ an enhancement at a level of 2-4\% (MOS1/pn) and 3-5\%
  (MOS2/pn). Tests suggest that the methods employed here are stable and
  robust.}
{The results presented here provide the most accurate cross-calibration of the
  effective areas of the XMM-Newton EPIC pn and MOS instruments to date. They
  suggest areas of further research where causes of the MOS-to-pn differences might be
  found, and allow the potential for corrections to and possible rectification
  of the EPIC cameras to be made in the future. }


   \maketitle

\section{Introduction}

XMM-Newton (Jansen \etal\ 2001), a cornerstone mission of ESA's Horizon 2000
science program, was designed as an X-ray observatory able to
spectroscopically study cosmic X-ray sources with a very large collecting area
in the 0.2$-$10\,keV band. This high throughput is achieved primarily through
the use of 3 highly nested Wolter type {\rm I} imaging telescopes. One of the
three co-aligned XMM-Newton X-ray telescopes has an unimpeded light path to
the primary focus, where the European Photon Imaging Camera (EPIC) pn camera
(Str\"{u}der \etal\ 2001) is positioned.  The two other telescopes have
EPIC-MOS cameras (Turner \etal\ 2001) at their primary foci, but receive only
approximately half of the incoming radiation, the remainder being diffracted
away by Reflection Grating Assemblies (RGAs) towards the secondary foci (where
the Reflection Grating Spectrometers (RGS; den Herder \etal\ 2001) are
situated).

The effective area of the EPIC instruments is defined here as the product of
the mirror effective area, the detector quantum efficiency and the filter
transmission, and is an extremely important quantity to have accurately
established. In detail, further complications can include: that the effective
area for the MOSs includes the RGA transmission factors, that one cannot
extract 100\% of the source counts, unless one uses unworkably large
extraction regions (i.e. Point Spread Function [PSF] issues), that
contamination of the mirror and/or detector may be present, that the full
detector efficiency can include other effects in addition to the quantum
efficiency, such as out-of-time events, and that the mirror effective area
includes vignetting, which, though minimised on-axis, is still important.
Furthermore, as the EPIC CCDs are sensitive to IR-UV light, a filter wheel
provides a choice of optical blocking filter, including thin, medium and thick
filters (plus open and closed). Usage of these filters alters the low-energy
X-ray transmission and requires extra specific detailed calibration.

Many attempts have been made to quantify more and more precisely the effective
area of the EPIC instruments (and other instruments, both on XMM-Newton and on
other X-ray observatories), and this has been one of the main drivers behind
IACHEC\footnote{http://web.mit.edu/iachec/} $-$ the International Astronomical
Consortium for High Energy Calibration, formed to provide standards for
high-energy calibration and to study in detail the cross-calibration between
different high-energy observatories.

This work describes a new attempt to establish the relative cross-calibration
accuracy of the on-axis effective areas of the XMM-Newton EPIC pn and MOS
instruments, using a sample of bright, isolated point sources, selected from
the 2XMM catalogue. The results provide the most accurate cross-calibration to
date of the EPIC pn and MOS effective areas. The structure of the paper is as
follows: Sec.~2 describes the sample selection and analysis. Sec.~3
describes the results obtained. In Sec.~4 we discuss these results and
further tests that were made, and in Sec.~5 we present our
conclusions. Unless otherwise stated, a reference to the term MOS1, MOS2 or pn
indicates that particular `whole telescope' system.

\section{Analysis}

\subsection{Sample Selection and Data Reduction}

Sources were selected from the entire 2XMM (2XMMi-DR3) catalogue (Watson
\etal\ 2008), as follows. The sources:

(i) Are identified within 2XMM as point-like (i.e. no extension), thus
maximizing the signal-to-noise ratio, simplifying and unifying the spectral
and effective area analysis, and maintaining consistency of extent across the
sample.

(ii) Have been observed in full-frame mode only. This is the most common EPIC
observing mode, and has the most accurate event energy calibration as the CCD
charge transfer inefficiency (CTI) models are best derived from full frame
data. Full-frame mode has 100\% livetime in both pn and MOS, and offers the
best background subtraction because of the active silicon relatively near any
on-axis source. Mode consistency is maintained across the sample.

(iii) Utilize either the thin or medium filter in all of MOS1, MOS2 \&
pn. These are the two most common EPIC filters employed. Consistency of filter
is maintained across the sample.

(iv) Have large numbers of (0.2$-$12\,keV) counts ($>$5000 for MOS1 or MOS2
and $>$15000 for pn). Though a larger number of fainter sources could be used
in addition to boost the statistics, much more screening would be necessary,
and background-modelling and source-to-background effective area issues would
become more significant.

(v) Have a countrate below the appropriate pile-up limits\footnote{XMM-Newton
  Users Handbook (v.2.10) :
  http://xmm.esac.esa. int/external/xmm\_user\_support/documentation/uhb/XMM\_UHB.html}
(0.70 ct/s [MOS full-frame], 6 ct/s [pn full-frame]). Pile-up distorts both
the spatial and spectral distributions of the photons away from the
well-understood non-piled-up case, and thus introduces extra unwanted problems
into the analysis. Although the potential `pile-up'-free sample could be
increased by using window mode data (with their shorter frame times),
full-frame mode data only is used (see above).

(vi) Are situated near on-axis, with EP\_OFFAX (the smallest value of the MOS1,
MOS2 or pn boresight-to-source distances) being $<$2\arcm. We are concerned here
with the effective area cross-calibration on-axis, where all the proposed EPIC
source targets are situated, and where the calibration is most mature (also
the vignetting is minimised). Later work, outside of the scope of the present
paper, hopes to look at the effective area cross-calibration off-axis, where
many additional issues come into play; CCD gaps, different CCDs, increased
vignetting, elongated PSFs, obscuration by the Reflection Grating Array (RGA)
etc.

(vii) Lie out of the plane of the Galaxy ($|b|$$>$15\,deg), where the Galactic
column density is generally lower than at low Galactic latitudes. This
minimses possible unknown effects due to absorption.

This gave rise to 87 sources. This is a very small number given the number of
sources in the 2XMM catalogue (over 350,000), and this is because the above
criteria applied are collectively extremely restrictive. For each of these
sources, the appropriate raw data - the Observation Data Files (ODFs) - for
each source were identified and obtained, and the standard SAS procedures
(`epchain' for pn, `emchain' for MOS) were run on these to create the standard
calibrated event lists. SAS version v12.0.0 and the then most up-to-date
public calibration files, created in late October 2012, have been used
throughout the analysis. Calibration is a continually evolving process, and as
such this analysis (like all others) provides only a snapshot of the situation
at the time of the analysis. It is worth noting that the latest SAS version
and calibration files at the time of writing include a correction in the MOS
effective area for an evolving contamination
layer\footnote{http://xmm2.esac.esa.int/docs/documents/CAL-SRN-0305-1-0.ps
  .gz}. This is modelled as a pure Carbon layer on the MOS detector and is
relatively thin, having an estimated depth at current epochs in 2013 of only
$0.015\,\mu$m and $0.04\,\mu$m on MOS1 and MOS2 respectively. This is less
than 20\% of the estimated thickness of the evolving Carbon contaminant known
to be on the RGS detector. There is no evidence as yet for a contaminant on
the pn. As this study deals with sources detected prior to October 2008, the
effect of the MOS contaminant model on this study is negligible.

The calibrated event lists were then filtered for periods of high background
(solar proton flares) via Good Time Interval (GTI) files. Single pattern
lightcurves were created in the 10$-$15\,keV band, with 100\,s binning, across
the full detectors (excluding MOS1/2 CCD1 and pn CCD4, which contain the
sources under study), and then GTIs were then defined as those bins containg
less than 130 (pn) or 40 (MOS) counts per bin. A common GTI (combining the
MOS1, MOS2 \& pn GTI files) was calculated, and this was then used to filter
the MOS1, MOS2 \& pn event files. 
Large-scale images around each source were created, and sources were rejected
in the following cases; 
(i) crowded fields, where neighbouring sources lay within the target source
  region, or where there was no appropriate source-free background region,
(ii) when the source had only a very short ($<$\,1ks) common GTI-filtered
exposure time,
(iii) where chip gaps or bad CCD columns were seen to lie too close to the
target source, 
(iv) where the source appeared to be extended (or to lie within extended
emission),
(v) there was a loss of an entire detector chip or quadrant. 
Cases (i) and (ii) accounted for the vast majority of the exclusions, and
this yielded a final sample of 46 sources, which are listed in Tab.\ref{tab1}.

\begin{table}
\caption{The final sample of 46 EPIC on-axis point sources. Tabulated values,
  taken from the 2XMM catalogue, are: 2XMM DetID, OBSID, revolution number,
  and pn, MOS1 and MOS2 (0.2-10\,keV) count rate. Also given is the common
  GTI-filtered exposure time.}
\label{tab1}
\begin{center} 
\begin{small}
\begin{tabular}{| r | r | r | c | c | c | r|}\hline
DetID & OBSID & Rev. & \multicolumn{3}{c|}{Count rate} & Exp.\\ 
 & & & \multicolumn{3}{c|}{(0.2-10\,keV; s$^{-1}$)} & GTI \\ \cline{4-6}
      &       &      & \hspace{-1mm}pn\hspace{-1mm} & \hspace{-1mm}MOS1\hspace{-1mm} & \hspace{-1mm}MOS2\hspace{-1mm} & (ks) \\ \hline

 18807 & 0303340101 & 1102 & 3.07 & 0.62 & 0.63 & 44.7\\
 22675 & 0084140501 & 0395 & 1.53 & 0.44 & 0.44 & 12.4\\
 22676 & 0084140101 & 0217 & 1.40 & 0.39 & 0.40 & 35.3\\
 29938 & 0200480101 & 0915 & 1.68 & 0.38 & 0.39 & 30.2\\
 31935 & 0300630301 & 1120 & 1.01 & 0.28 & 0.27 & 17.6\\
 37130 & 0106860101 & 0157 & 0.92 & 0.31 & 0.30 & 23.4\\
 46636 & 0086360401 & 0230 & 1.11 & 0.31 & 0.31 & 26.9\\
 46637 & 0086360301 & 0230 & 1.66 & 0.43 & 0.45 & 54.5\\
 48184 & 0101441501 & 0410 & 1.68 & 0.42 & 0.42 & 37.1\\
 48185 & 0101440601 & 0138 & 2.33 & 0.57 & 0.60 & 34.7\\
 70141 & 0153250101 & 0615 & 1.90 & 0.38 & 0.38 & 54.7\\
 97435 & 0300910301 & 1068 & 1.71 & 0.38 & 0.39 & 13.0\\
100165 & 0201290301 & 0813 & 0.99 & 0.29 & 0.29 & 18.2\\
102212 & 0112521001 & 0428 & 2.18 & 0.68 & 0.69 & 7.9\\
112114 & 0151390101 & 0630 & 0.71 & 0.21 & 0.21 & 46.3\\
127689 & 0112850201 & 0469 & 1.87 & 0.60 & 0.60 & 16.5\\
127982 & 0112880101 & 0565 & 2.02 & 0.50 & 0.52 & 28.1\\
137187 & 0205010101 & 0816 & 1.74 & 0.47 & 0.48 & 26.3\\
141829 & 0306630201 & 1103 & 1.34 & 0.32 & 0.32 & 91.4\\
142557 & 0124900101 & 0082 & 2.17 & 0.60 & 0.61 & 29.7\\
152354 & 0204040301 & 0841 & 1.47 & 0.39 & 0.39 & 56.4\\
152355 & 0204040101 & 0823 & 1.95 & 0.52 & 0.52 & 75.6\\
154252 & 0300240501 & 1129 & 1.17 & 0.39 & 0.40 & 25.8\\
158032 & 0206580101 & 0939 & 0.64 & 0.17 & 0.17 & 36.9\\
163564 & 0147670201 & 0576 & 1.59 & 0.48 & 0.47 & 12.1\\
163784 & 0152940101 & 0660 & 1.29 & 0.43 & 0.43 & 39.4\\
167394 & 0205390301 & 0831 & 0.73 & 0.18 & 0.19 & 49.1\\
171106 & 0092850501 & 0373 & 1.09 & 0.28 & 0.28 & 38.9\\
176741 & 0207130401 & 0948 & 2.08 & 0.54 & 0.55 & 11.7\\
178470 & 0067750101 & 0311 & 1.23 & 0.31 & 0.32 & 23.6\\
182637 & 0100240701 & 0127 & 2.15 & 0.61 & 0.65 & 15.5\\
191302 & 0056340201 & 0420 & 2.18 & 0.64 & 0.66 & 12.4\\
219702 & 0146390101 & 0604 & 1.66 & 0.59 & 0.57 & 17.3\\
234746 & 0147920601 & 0539 & 2.12 & 0.67 & 0.69 & 12.1\\
248779 & 0402560901 & 1291 & 0.71 & 0.22 & 0.22 & 48.0\\
255122 & 0405090101 & 1255 & 0.88 & 0.28 & 0.28 & 96.1\\
272445 & 0304320201 & 1017 & 1.52 & 0.41 & 0.41 & 68.5\\
272446 & 0304320301 & 1016 & 1.40 & 0.37 & 0.37 & 36.1\\
272447 & 0304320801 & 1189 & 1.98 & 0.52 & 0.52 & 37.8\\
273965 & 0405690501 & 1275 & 0.85 & 0.23 & 0.24 & 25.8\\
273967 & 0405690201 & 1272 & 1.15 & 0.31 & 0.32 & 36.1\\
296606 & 0510181701 & 1528 & 1.59 & 0.51 & 0.52 & 44.6\\
310154 & 0502220301 & 1441 & 0.31 & 0.10 & 0.09 & 69.1\\
319432 & 0555020201 & 1558 & 2.16 & 0.60 & 0.61 & 24.9\\
332369 & 0555650201 & 1598 & 0.64 & 0.18 & 0.18 & 96.4\\
332370 & 0555650301 & 1599 & 0.63 & 0.17 & 0.17 & 92.8\\

\hline
\end{tabular}
\end{small}
\end{center}
\end{table}

\subsection{Data Analysis and Spectral Reduction}

For each source and for each instrument, unbinned source spectra were
extracted from a complete circle of radius 40\arcs, centred on the 2XMM source
position. A background spectrum was extracted from a complete 90$-$180\arcs
annulus, again centred on the 2XMM source position. Cases where the background
region appeared contaminated with other sources were removed in the above
screening stage, thus keeping both the source (pure circle) and background
(pure annulus) extraction regions as clean and as simple as possible. For each
source, the three EPIC instruments each cover the same source and background
regions on the sky. The estimated source contribution to the background is
equally low ($<$5\%) for all three cameras, and this total background only
accounts for $\approx$2\% of the extracted total source flux. The spectral
extraction criteria that is
recommended\footnote{http://xmm.esac.esa.int/external/xmm\_user\_support/documentatio
  n/sas\_usg/USG/} to the general XMM-Newton user were employed: PATTERN no
greater than 12 for MOS, no greater than 4 for pn, flag selections of
\#XMMEA\_EM for MOS, and FLAG==0 for pn, and using spectralbinsize=5 in
evselect. Standard energy response matrices (RMFs) and effective area files
(ARFs) were created for each spectrum via the standard SAS tools.

Then for each instrument, the source spectral counts $C$ were stacked
together, using various ftools\footnote{https://heasarc.gsfc.nasa.gov/ftools/}
(and exposure-weighting the BACKSCAL values):

$$
C_{i} = \sum_{j} C_{ij}
$$

where $i$ refers to the $i$\,$^{\rm th}$ spectral bin and $j$ refers to the
$j$\,$^{\rm th}$ source. The background spectral counts $B$ were stacked
together in the same manner (exposure-weighting the BACKSCAL values).

$$
B_{i} = \sum_{j} B_{ij}
$$

Slightly different source and background BACKSCAL values are seen from source
to source within each detector, as different (and time-dependent) bad pixels
and columns reduce the `good' detector areas. An average exposure-weighted ARF
was calculated, summing the effective areas ($EA$) in the same way, using
ftools:

$$ 
EA_{i} = \sum_{j} EA_{ij}
$$

An average exposure-weighted RMF was calculated using the ftool addrmf, while
ensuring that the sum of the individual exposure-weights was unity. The final
output for each instrument (MOS1, MOS2 \& pn) is then: one source spectrum,
one background spectrum, one ARF \& one RMF.

The spectral analysis that follows is confined to energies above 0.5\,keV
where relative uncertainties in the instruments are dominated by the
components of the effective area calibration. At lower energies, the RMF of
both pn and MOS becomes strongly non-Gaussian in shape and the interpretation
of residuals in the cross-calibration becomes increasingly influenced by
uncertainties in the RMF calibration.

\subsection{Spectral Analysis}

To examine the relative cross-calibration accuracy of the effective areas of
the EPIC instruments, we applied the method of Longinotti et al. (2008), in a
similar manner to that performed in Kettula et al. (2013).  This method
involves a choice of a reference instrument, against which the other
instruments are compared.  As it is not known which, if any, of the
instruments has a perfectly calibrated effective area, the choice of the
reference instrument is arbitrary and the comparison is a relative one. The
pn detector was chosen as the reference instrument, reasons being that
the more sensitive pn mainly `drives' any all-EPIC fit, and that pn appears to
be extremely stable; pn
flux measurements below 1\,keV are stable within 3\% (at the 3$\sigma$ level)
over the whole of XMM-Newton's operational life (Sartore \etal, 2012), though
this is probably still a very conservative upper limit (Pollock
\etal, in preparation), and pn is likely in fact to be much more stable. 

\begin{figure*}
\centering
\includegraphics[width=16cm]{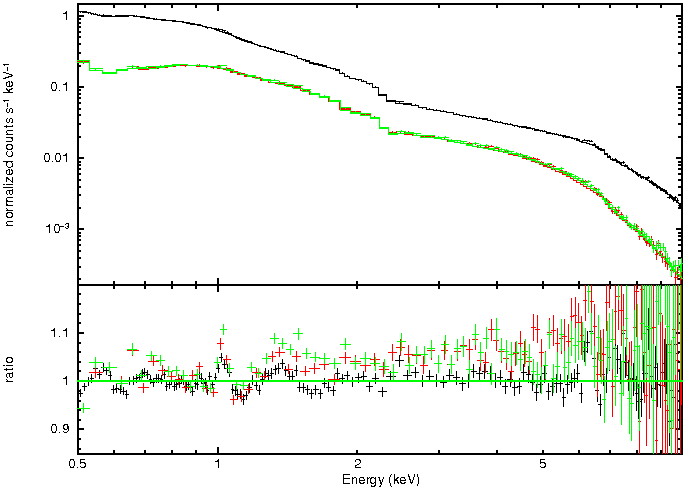}
\caption{Spectral fitting of the full stacked data from the sample of 46
  sources. A phenomenological spectral model is fit closely to the pn data
  (black points). This model (for details, see text) is then convolved with
  the instrument responses of MOS1 (red) and MOS2 (green). Data and model are
  shown in the upper panel, and the residuals are shown in the lower
  panel. Ratioing the residuals reveals how MOS1 and MOS2 vary with respect to
  pn.}
\label{fig_spec}
\end{figure*}

How MOS1 and MOS2 vary with respect to pn can then be inspected.  This method
$-$ firstly stacking all the data together, and then performing spectral
fitting on these stacked data $-$ we will later refer to as a `stack \& fit'
method. For two instruments, here pn and MOS, we begin with the data
(data\,$_{\rm pn,MOS}$) and the responses (resp\,$_{\rm pn,MOS}$), defined
here as the energy response matrix (RMF) multiplied by the effective area
(ARF) of the instruments. The pn is used as the reference instrument (pn) and
spectral analysis is performed on the pn data to fit a reference model
(model\,$_{\rm pn}$). Several multi-component phenomenological models were
constructed to closely fit the pn data to varying degrees of precision. The
details of the phenomenological models are actually not important in this
study, as usage of models of different complexity $-$ using models of varying
`goodness' $-$ produced essentially no changes in the final result, the $\alpha$
ratio (see below). Typically though, the models included some absorption, one
or two power laws, a small number of Gaussians, and sometimes an edge. A model
that closely fits the pn data is shown (with residuals) in
Fig.~\ref{fig_spec}, and is as follows: wabs
($n_{H}=1.84\times10^{21}$\,cm$^{-2}$) $\times$ [power ($\Gamma=4.09$) + power
  ($\Gamma=1.43$) + gauss ($E=0.59$\,keV) + gauss ($E=0.88$\,keV) + gauss
  ($E=5.03$\,keV) ] $\times$ edge ($E=6.85$\,keV).  The reduced chi-squared
for this model fit to the pn data is 1.19 for 1888 degrees of freedom.

To compare the two instruments, we ensure that the data (data\,$_{\rm
  pn,MOS}$) are of equal binning. We note here the minor caveat that the
binning is in PI space, which may not necessarily correspond to exactly the
same photon energy ranges, due to the slightly different redistributions in
MOS and pn, however this effect is negligible in our case, as the energy bins
of the stacked residual spectra shown in this paper are larger than, or at
least comparable to the instrumental energy resolution of the EPIC cameras. We
convolve the reference model (model\,$_{\rm pn}$) with the instrument
responses of MOS1 and MOS2 to produce pn-based model MOS predictions
(model\,$_{\rm MOS\,|\,pn}$). We then divide the MOS data by these model
predictions to obtain residuals over the pn-based prediction (resid\,$_{\rm
  MOS}$) The model MOS predictions and residuals are also shown in
Fig.~\ref{fig_spec}.

Lastly, to obtain the comparison between MOS and pn, we divide the above
residuals of MOS with the pn residuals, to remove effects of possible
calibration uncertainties of pn, obtaining a ratio $\alpha$ 
$= {\rm resid\,}_{\rm MOS} /  {\rm resid\,}_{\rm pn}$

\begin{equation}\label{eq_r}
\alpha = \frac{{\rm data\,}_{\rm MOS}}{{\rm model\,}_{\rm MOS\,|\,pn} \otimes {\rm resp\,}_{\rm MOS} } 
    \times 
    \frac{{\rm model\,}_{\rm pn} \otimes {\rm resp\,}_{\rm pn} }{{\rm data\,}_{\rm pn}}
\end{equation}
\vspace{2mm}

The ratio $\alpha$ is then a useful measure of the effective area cross-calibration,
since the value for each energy bin would be unity if the cross-calibration of
the pn and MOS effective areas were consistent.

In a more formal analysis, along the lines of e.g. Marshall \etal\ (2004), we
can start from the formal equation for the count spectrum for one source $j$
in the sample for instrument $xx$ (where $xx$=MOS or pn):

\begin{equation}
C^{xx}_{ij} = T_{j} \int A^{xx}_{j}(E) R^{xx}_{ij}(E) n_{j}(E) dE + a^{xx}_{j} B_{ij}
\end{equation}

where $T_{j}$ is the exposure time of the $j$th source, $A^{xx}_{j}$ is the
effective area for source $j$ for instrument $xx$, $a^{xx}$ refers to the
BACKSCAL parameter for source $j$ for instrument $xx$, $B_{ij}$ is the
background count spectrum for source $j$, $n_{j}(E)$ is the photon spectrum of
source $j$, and $R^{xx}_{ij}(E)$ is the RMF for source $j$, giving the
fraction of events going into instrument $xx$'s PI bin $i$ that originate at
energy $E$.

The MOS-to-pn ratio of source counts for the stacked spectra is therefore:

\[
\frac{\displaystyle\sum_{j} C^{MOS}_{ij} - \displaystyle\sum_{j} a^{MOS}_{j} B^{MOS}_{ij}}
     {\displaystyle\sum_{j} C^{pn}_{ij}  - \displaystyle\sum_{j} a^{pn}_{j}  B^{pn}_{ij}} 
\hspace{5mm}
=
\]
\begin{equation}
\hspace{25mm}
\frac{\displaystyle\sum_{j} T_{j} \int n_{j}(E) \widehat{A}^{MOS}_{j}(E) R^{MOS}_{ij}(E) dE}
     {\displaystyle\sum_{j} T_{j} \int n_{j}(E) \widehat{A}^{pn}_{j}(E) R^{pn}_{ij}(E) dE}
\end {equation}

where $\widehat{A}^{MOS}$ and $\widehat{A}^{pn}$ are the {\em true} MOS and
pn effective areas. 

We can then define a parameter $\alpha^{\prime}$ relating the ratio of the effective
areas currently assumed to be correct to the {\em true} effective areas:

\begin{equation}
\alpha^{\prime} = \frac { A^{MOS} / A^{pn} } { \widehat{A}^{MOS} / \widehat{A}^{pn} }
\hspace{5mm}
= 
\frac{A^{MOS}}{\widehat{A}^{MOS}} \times \frac{\widehat{A}^{pn}}{A^{pn}} 
\end{equation}

This $\alpha^{\prime}$ parameter is equal to the $\alpha$ ratio above in
equation~\ref{eq_r}, and we have:

\[
\alpha = 
\frac{\displaystyle\sum_{j} C^{pn}_{ij}  - \sum_{j} a^{pn}_{j}  B^{pn}_{ij}}
     {\displaystyle\sum_{j} C^{MOS}_{ij} - \sum_{j} a^{MOS}_{j} B^{MOS}_{ij}}
\hspace{5mm}
\times
\]
\begin{equation}
\hspace{25mm}
\frac{\displaystyle\sum_{j} T_{j} \int n_{j}(E) A^{MOS}_{j}(E) R^{MOS}_{ij}(E) dE}
     {\displaystyle\sum_{j} T_{j} \int n_{j}(E) A^{pn}_{j}(E) R^{pn}_{ij}(E) dE}
\end{equation}

This is true as $\int R^{xx}_{ij}(E)dE$ is unity for both the MOS and pn
instruments. Also as long as $n_j(E)$ and $A_j (E)$ only vary slowly over a
spectral channel $-$ this in principle does not hold at energies where the
effective area has the steepest gradient, {\it i.e.} close to the Si
($\simeq$1.8\,keV) and Au ($\simeq$2.2\,keV) edges. This may introduce
features in the stacked spectrum. While one indeed sees structures in the
residuals of the stacked spectrum against its best-fit model, the amplitude of
these features is comparable to that observed when fitting individual
spectra. They can be reconnected to imperfect calibration of the MOS Quantum
Efficiency, and EPIC effective areas, respectively.

\section{Results}

Fig.~\ref{fig_spec} shows (by design) a very good fit to the pn data, and a
less good fit to the MOS data $-$ the MOS1 and MOS2 data points appear to lie
slightly above the model across much of the energy range $-$ this is examined
in more detail below. The method described here has removed or minimized as
many sources of error or uncertainty as possible. Source variability cannot be
the cause of the MOS-to-pn differences, as common GTIs across all the instruments
have been used for each of the sources. All the sources are below the pile-up
limit in all the instruments, so pile-up is not an issue. Consequently, it has
therefore been possible to use a large, full circular region for the spectral
extraction, such that any PSF inaccuracies have been minimized
(i.e. importantly we have not had to excise any piled-up central
core). Furthermore, the problem that all the sources may be spectrally rather
different, some having perhaps complex spectra, is overcome by stacking all
the spectra together into a single spectrum, and fitting this
phenomenologically. We discuss this issue further in Sec.~\ref{sec_disc}.

\begin{figure}
\centering
\includegraphics[width=8.8cm]{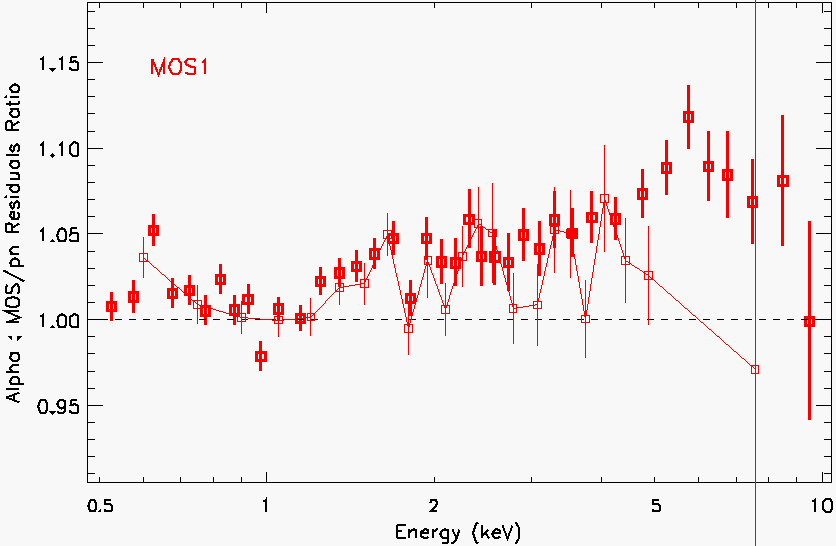}
\includegraphics[width=8.8cm]{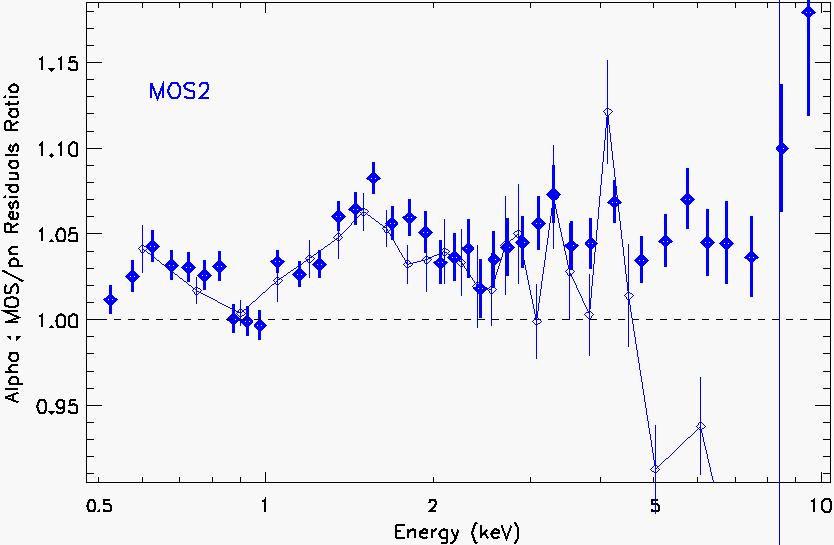}
\caption{How MOS1 (top panel) and MOS2 (bottom panel) vary with respect to
  pn. $\alpha$ ratios for MOS1/pn (red) and for MOS2/pn (blue) are plotted
  against energy (bold points). The data have been adaptively binned to follow
  the statistics. The faint points, connected by a thin line, show the
    results obtained using the `fit \& stack' (see Sec.~\ref{sec_disc})
    analysis (the last MOS2/pn point is at $\alpha$$\approx$0.75).}
\label{fig_m12}
\end{figure}

Once we have the pn, MOS1 and MOS2 residuals, we are then able to ratio these
residuals to obtain $\alpha$, and show how MOS1 and MOS2 vary with respect to
pn. This ratio, described above, removes any differences between the pn data
and the pn model prediction, and is seen to be insensitive to the general
quality of the pn fit (so long that it is adequate $-$ i.e. regardless of the
quality of the fit, so long that a by-eye inspection of the residuals
indicated that no spectral feature was still unaccounted for, then no change
in the residual ratio $\alpha$ is observed). These $\alpha$ ratios for MOS1/pn and for
MOS2/pn are shown as the bold points in Fig.~\ref{fig_m12}. Here, adaptive
binning of the results has been used to follow the quality of the statistics
across the energy range. At low energies ($<$1\,keV), small (0.05\,keV) bins
have been used, as there are many counts. At high energies ($>$7\,keV), large
(1.0\,keV) bins have been used, as there are fewer counts.

To first order, the MOS1 and MOS2 plots appear very similar. Generally, the
MOS/pn $\alpha$ ratio is, at low energy ($<$1\,keV), flat and low
($\alpha\approx1.00-1.03$). This is the energy range over which MOS and pn agree
best. Then over the medium energy range, it rises, to flatten out again beyond
3\,keV at a higher (close to) energy-independent value
($\alpha\approx1.05-1.08$). On top of this function, where the growth is
slowest at the beginning and the end of the function, there is a `bump' at
1$-$2\,keV, which we hereafter call the `silicon bump', and which may be
larger in MOS2 than in MOS1. There are perhaps in addition other small
differences between MOS1 and MOS2 that are close in size to the level of the
statistical uncertainties; at high energies MOS1/pn may be slightly higher
than MOS2/pn, whereas at low energies the reverse may be the case.

\section{Discussion}
\label{sec_disc}

There have been a number of attempts to accurately quantify the effective area
- the product of the mirror effective area, the detector quantum efficiency
and the filter transmission - of not only the EPIC instruments, but of other
X-ray instruments also. The various strategies have certain advantages and
disadvantages. Kettula \etal\ (2013) for instance, studied a sample of
clusters of galaxies with the XMM-Newton/EPIC, Chandra/ACIS, BeppoSAX/MECS and
Suzaku/XIS instruments. The advantages of using clusters of galaxies is that
they are constant and unvarying and they are spectrally quite simple. However
clusters are also extended and diffuse and this complicates the analysis,
especially with regard to modelling the background and PSF issues. Very bright
point sources, e.g. radio loud AGN, have been used for cross-calibration
purposes (e.g. Stuhlinger \etal\ 2010). Here the advantages include the fact
that the sources are very bright and, unlike the clusters,
point-like. However, these sources are often too bright and consequently
suffer from pile-up. Dealing with the pile-up, usually by excising the core of
the emission (a non-optimum method given the azimuthal profile of the
emission), can introduce errors if the PSF is not extremely accurately
calibrated. Also these sources are variable. This problem can be circumvented
by defining common GTIs to cover all the instruments in question. In the
present paper, we have used various 2XMM point sources that are less bright
and are therefore not piled-up. We do not need to excise the core and a full
circular extraction region can be used, minimising any PSF issues. Another
point is that the selection criteria employed only yielded a few tens of
sources from the initial 2XMM catalogue of 350,000. This is an extremely small
number, the criteria being extremely restrictive, and indicates why such a
study as this has not been possible before now $-$ many years of mission time
have been needed in order to accumulate enough suitable sources. These sources
may be variable (many of the 46 sources are AGN), but common GTIs have been
used to deal with this. One remaining issue with the 2XMM sample is that the
sources may be spectrally complex and/or different from one another. We
discuss this issue in the following paragraphs.

In the analysis of Kettula \etal\ (2013), the residuals ratio analysis was
performed on each source in their sample separately, and then the median and
median absolute deviation of their $\alpha$ values were calculated for each
energy channel for their sample. This represents more of a `fit \& stack'
analysis, as opposed to the `stack \& fit' analysis presented in this
paper. To test the validity of both methods, a `fit \& stack' analysis was
performed on our sample of 46 sources. This involved the spectral fitting of
each of the 46 pn spectra separately, then convolving each model with the
individual MOS instrument responses, to obtain 46 separate residual ratio
datasets, from which the median and median absolute deviations could be
calculated. These individual spectral fits were performed using $\chi^2$
goodness-of-fit tests, after standard spectral grouping using
`specgroup'. Only those sources where the reduced $\chi^2$ was lower than 1.5
were retained in the final median analysis. We note that likelihood-based
tests may have yielded a better agreement.

The results for the `fit \& stack' method are shown as the faint points
connected by thin lines in Fig.~\ref{fig_m12}. It is seen that the agreement
between the `stack \& fit' and the `fit \& stack' methods on the sample of 46
sources is very good in both MOS1 and MOS2 below 4\,keV. Above 7\,keV for
MOS1, and above 5\,keV for MOS2, a deficit in the $\alpha$ ratio is seen for
the `fit \& stack' method, which is not seen in the `stack \& fit'
results. Investigating further, we tentatively attribute the deficit to a
background subtraction problem involving the error bars on the negative
spectral bins that can sometimes occur in the individual source spectra, but
not in the (high statistic) stacked spectrum. We conclude that the `stack \&
fit' method yields consistent results to the `fit \& stack' method, and
probably gives better results over energy ranges where there are many negative
background-subtracted channels (due to statistical fluctuations of the
background subtraction). If anything, the `stack \& fit' method used here
appears to be the more robust of the two.

An important epoch-dependent effect is the low-energy on-axis `patch' in both
MOS instruments (Read \etal\ 2005). Here, the redistribution function is seen
to evolve with time such that an increasing fraction of X-rays suffer
incomplete charge collection. The effect is increasingly stronger towards
lower energies, and appears to be related to the total X-ray radiation dose
received by a pixel which will clearly be higher at the centre of the
detector. Sources at MOS off-axis angles greater than ∼2\arcm\ are essentially
immune to this effect. For MOS on-axis sources, photons from the energy band
0.2-0.5\,keV can be redistributed in energy below the detection threshold and
lost, while photons from energy band 0.5-1.0\,keV can be redistributed into
the 0.2-0.5\,keV energy band. The effect is negligible at energies above
∼1\,keV.  The `patch effect' has been extensively studied, and is well
calibrated. All but 4 of the 46 sources are seen to lie on the MOS patch. As a
sanity check, we looked at the results obtained with the full stacked data,
but using single RMFs from the beginning, the middle, and the end of the
mission time period considered (revolutions $<$1800). The only effects seen
were (as expected) those at low-energy due to the evolving patch and were
entirely consistent with the known (and calibrated) instrumental
evolution. Usage of the beginning (end) RMF introduces a downwards (upwards)
offset of 5\% in the $\alpha$ ratio at 0.5\,keV, for both MOS1/pn and MOS2/pn,
when compared with using the correct exposure-weighted average of the
RMFs. This offset decreases to zero at 1.3\,keV and remains at zero out to
10\,keV. The `stacked' RMF that we have employed, an exposure-weighted average
of the RMFs constructed for each separate source, is seen to behave entirely
correctly, and as expected when compared with the single RMF cases and
factoring in the epoch- and exposure-distrutions of the contributing files.

It is hard to resolve the cross-calibration discrepancies between the EPIC
cameras reported here just by changing the quantum efficiency calibration,
without violating the ground-based measurements (see Turner \etal\ 2001,
Saxton \& Sembay 2004). However, the overall shape of the stacked residuals
indicates that a change in the quantum efficiency could contribute to
alleviating the problem. A change in the energy-dependent quantum efficiency
of the MOS detectors, say, which has an intrinsically larger silicon edge than
the pn, could remove at least part of the `silicon bump' and part of the
`slow-faster-slow' energy-dependence of the MOS/pn beaviour. Furthermore, a
mis-calibration of the silicon fluorescence peak may be a cause of part of the
`silicon bump'. Further analysis may help in understanding which other
calibration element(s) might be involved, and might require further detailed
investigation.

\section{Conclusions}
\label{sec_conc}

We have examined the accuracy of the relative cross-calibration of the on-axis
effective areas of the XMM-Newton EPIC pn and MOS instruments. A sample of 46
bright, high-count, non-piled-up, isolated, on-axis point sources was selected
from the 2XMM catalogue. After flare- and background-cleaning, and applying
common GTI filtering, source and background spectra extracted from the 46
sources were stacked together, and the individual response files were averaged
in an exposure-weighted manner. Spectral fitting was applied, and the MOS and
pn model residuals were examined in order to characterize the EPIC MOS-to-pn
inter-calibration. It was seen that the MOS1-to-pn and MOS2-to-pn results are
broadly very similar, with the cameras showing the closest agreement below
1\,keV, with MOS excesses over pn of 0-2\% (MOS1/pn) and 0-3\%
(MOS2/pn). Above 3\,keV, the MOS/pn ratio is consistent with an
energy-independent (or only mildly increasing) ratio, with MOS excesses of
7-8\% (MOS1/pn) and 5-8\% (MOS2/pn). In addition to this, between 1-2\,keV
there is a further excess $-$ a `silicon bump' $-$ at a level of 2-4\%
(MOS1/pn) and 3-5\% (MOS2/pn). Tests reveal that the `stack \& fit' methods 
employed here appear to be stable and robust, and the results presented here
provide the most accurate to date cross-calibration of the on-axis effective
areas of the XMM-Newton EPIC pn and MOS instruments. Areas of research where
possible causes of the MOS-to-pn mismatches might be found are suggested by
the analysis, and we note the potential for future corrections to and possible
rectification of the EPIC MOS and pn cameras to be made.

\begin{acknowledgements}

The XMM-Newton project is an ESA Science Mission with instruments and
contributions directly funded by ESA Member States and the USA (NASA). We
thank members of the EPIC cross-calibration working group $-$ Jukka
Nevalainen, Martin Stuhlinger, Frank Haberl, Silvano Molendi, Richard Saxton
\& Michael Smith $-$ for helpful discussions.  We also thank the referee
Herman Marshall for very useful comments and discussions which have improved
the paper. AMR and SS acknowledge the support of STFC/UKSA/ESA funding.

\end{acknowledgements}


\end{document}